\documentclass[a4paper,fleqn]{cas-dc2}
\usepackage{cas-dfrws}
\renewcommand{\dfrwsconference}{Digital Forensics Research Conference USA (DFRWS USA), 2023}
\pdfoutput=1

\usepackage{stix}

\usepackage{graphicx}
\usepackage{amsmath}
\usepackage{amsthm}
\usepackage{lipsum}
\usepackage[dvipsnames]{xcolor}
\usepackage{mathtools}
\usepackage[colorlinks]{hyperref}
\hypersetup{
    linkcolor=CornflowerBlue,
    citecolor=CornflowerBlue,
    urlcolor=cyan
}
\usepackage[hyphenbreaks]{breakurl}
\usepackage{array,multirow}
\usepackage{caption, subcaption}
\usepackage[binary-units]{siunitx}
\usepackage{enumitem}
\usepackage{listings}
\lstdefinestyle{basic}
{
    frame=tb,
    rulecolor=\color{lightgray},
    backgroundcolor=\color{black!1!},
    basicstyle=\footnotesize\color{black}\ttfamily,
    commentstyle=\color{commentsColor}\textit,
    breaklines=true,
    aboveskip=0.2cm,
    belowskip=0.0cm,
    escapeinside={\%*}{*)}
}
\newcommand{\id}[1]{\textit{#1}}

\usepackage{tikz}
\usetikzlibrary{shapes.multipart}
\usetikzlibrary{arrows,calc,shapes,decorations.pathreplacing,matrix,calligraphy}
\usetikzlibrary{decorations.markings}
\usetikzlibrary{backgrounds,positioning,automata,arrows.meta}
\usetikzlibrary{tikzmark}

\usepackage{pgfplots}
\pgfplotsset{compat=newest}
\usepackage[framemethod=TikZ]{mdframed}
\mdfdefinestyle{TikzFrame}{%
    roundcorner=2pt,
    innertopmargin=4pt,
    innerbottommargin=4pt,
    innerrightmargin=0pt,
    innerleftmargin=4pt,
    leftmargin = 4pt,
    rightmargin = 4pt
}

\usepackage[numbers,authoryear,longnamesfirst]{natbib}
\begin{document}

\title[mode=title]{Systematic Evaluation of Forensic Data Acquisition using Smartphone Local Backup}
\shorttitle{Evaluation of Smartphone Local Backup}

\shortauthors{Geus et~al.}

\author[1]{Julian Geus}[orcid=0009-0001-8270-1964]
\cormark[1]
\ead{julian.geus@fau.de}
\credit{Conceptualization, Methodology, Investigation, Writing - Original Draft, Writing - Review and Editing}

\author[1]{Jenny Ottmann}[orcid=0000-0003-1090-0566]
\cormark[1]
\ead{jenny.ottmann@fau.de}
\credit{Conceptualization, Methodology, Investigation, Writing - Original Draft, Writing - Review and Editing}

\author[1]{Felix Freiling}[orcid=0000-0002-8279-8401]
\cormark[1]
\ead{felix.freiling@fau.de}
\credit{Conceptualization, Methodology, Investigation, Writing - Original Draft, Writing - Review and Editing, Supervision}

\address[1]{Department of Computer Science,
  Friedrich-Alexander-Universit\"at Erlangen-N\"urnberg (FAU),
  Erlangen, Germany}

\cortext[1]{Corresponding authors.}

\nonumnote{Copyright remains with the authors.}

\begin{abstract}
  Due to the increasing security standards of modern smartphones,
  forensic data acquisition from such devices is a growing
  challenge. One rather generic way to access data on smartphones in
  practice is to use the local backup mechanism offered by the mobile
  operating systems. We study the suitability of such mechanisms for
  forensic data acquisition by performing a thorough evaluation of
  iOS's and Android's local backup mechanisms on two mobile devices.
  Based on a systematic
  and generic evaluation procedure comparing the contents of local
  backup to the original storage, we show that in our exemplary practical evaluations,
  in most cases (but not all) local backup actually yields a correct copy of the original
  data from storage. Our study also highlights corner cases, such as
  database files with pending changes, that need to be considered when
  assessing the integrity and authenticity of evidence acquired
  through local backup.
\end{abstract}


\maketitle

\section{Introduction}
Despite many advances over the years
\citep{DBLP:journals/dt/FreilingGLMP18}, forensic data acquisition is a
tough challenge, especially for smartphones, where security features
are making attempts of unauthorized data collection from the devices
themselves increasingly futile. Many mobile devices, however, are
heavily integrated into networked ecosystems so that the devices
themselves are usually not the only storage location of interesting
data. Especially smartphones today are in many cases merely the cache
of data that is in fact stored on remote servers. Furthermore, cloud
storage is commonly used as the location of backup data at the request
of users. This has sparked an interesting stream of research in
\emph{cloud forensics}
\citep{DBLP:journals/di/RoussevABMS16,DBLP:journals/csur/ManralSCCG20}.

Tool manufacturers have, however, cleverly observed that backup data
can be accessed even without access to cloud storage. The trick is to
instruct smartphones to perform a \emph{local} backup to an attached
laptop and acquire data from there. Judging from multiple blog posts,
this method appears to be commonly used by established ``black box''
tools for forensic data acquisition~\citep{oxy_backup, cel_backup,
magnet_backup, cel_backup_ios}.
But despite the fact that local backup appears to be one of the few
device and manufacturer independent methods for data acquisition that
does not need elevated privileges, we are not aware of any related
work that assesses the reliability of such acquisition methods and the
quality of data which they acquire.

The fact that nobody currently appears to question the suitability of
local backup as a sound evidence acquisition method is worrisome since
a negative answer would not only affect future cases but also old
ones. Consider, for example, a case from the past in which decisive
evidence had been acquired from a mobile phone using the software from
a well-known forensic service provider. If it turned out that that
software was using local backup and that local backup acquisition is
commonly incomplete, unreliable or, even worse, often leads to data
alteration, the validity of the evidence may be challenged, which
could entail a reconsideration of the case.

One reason why local backup methods have not yet been evaluated could
be that it is not exactly clear how to do it for two main
reasons. Firstly, with Android and iOS there are two large families of
backup methods with (at least for Android) multiple variants that are
hard to compare. This matter is further complicated by the fact that
the backup process is usually app oriented, which leads to even more
variance. Secondly, backups regularly run concurrently to the OS and
applications which makes it hard to define a ground truth to which any
backup data can be compared.

\subsection{Contributions}

The main contribution of this paper is a generic yet practical
evaluation procedure to assess the reliability and data quality of
data acquired through local backup. We demonstrate its usefulness by
executing it on the two dominating smartphone platforms, Google's
Android and Apple's iOS.  More precisely, we evaluate what data can
and cannot be retrieved from using the local backup service of the
platform and to what degree the data retrieved from the backup
corresponds to the data that was originally stored on the device.  Our
evaluation procedure can handle different types of backup, namely
content-based as well as file-based backup, and allows for easy
repeatability of the evaluation for future versions of operating
systems and apps.

More concretely, we executed our procedure on an instance of Android's
full local backup and specific package data backup with
app-downgrading for various applications, as well as iOS's encrypted
and unencrypted local backups.
We limit the evaluation to one Android and one iOS device, because the
differences between the OS versions are unlikely to be significant, but
the practical execution entails considerable efforts.
For every case, we executed a total of 20 evaluation iterations to
provide reliable results and to locate outliers.  The results for
Android showed that in the large majority of cases, the acquired data
was the same as the source data on the device. The cause of the cases
where files showed alterations could most likely be accounted to
concurrent background processes since those occurred at random.
The results for iOS were
similar, but most database files showed alterations compared to their
source counterpart. Further investigation showed that this was due to
the merging of uncommitted database changes during the backup
process. These altered files accounted for over 10\si{\percent} of the
data in both encrypted and unencrypted backup runs, which both had
around 700 files included in the backup.

For our imaginary legal case, we therefore can attest the forensic
suitability of the evidence with the aforementioned limitations.

In summary the contributions of this paper are the following:
\begin{itemize}

\item An in-depth categorization of different backup types and an analysis
  of the amount of included data.

\item The development of a generic evaluation procedure which takes
  the type of data into account.

\item Practical backup acquisition evaluations using an Apple iPhone
  and an Android device.

\item The creation of comprehensive backup datasets, which we make
  accessible to the community \citep{grajeda2017availability} upon
  request.\footnote{Due to the inclusion of sensitive data like
    geolocation information, we are not able to make our datasets
    available for unauthenticated public download. Interested
    researchers should contact the authors to get access to the
    datasets.}

\end{itemize}

\subsection{Related Work}

Analyses of backup data acquisition are often described in various
blog posts from forensic service providers e.g.~\cite{sal_adb,
  far_ios}. For the app-downgrading process on Android devices,
\cite{oxy_down, oxy_ad_limit} posted information about the procedure
and its caveats.  As mentioned above, to the best of our knowledge, no
comparable systematic and scientific evaluation attempts for data
acquisition using local backups of smartphones have been carried out
so far.

Related work in a broader sense was done
by~\cite{chang2015jailbroken}, who created an overview of extractable
data on jailbroken and non-jailbroken iPhones and evaluated the impact
of jailbreaking on the acquired backup data. They concluded, that the
jailbreak does not seem to have an impact on the data.
Similarly,~\cite{hassan2017investigation} analyzed the impact of
rooting on the acquired data of an Android device, with the conclusion
that no changes could be observed. These studies were carried out on
outdated OS versions using outdated jailbreaking and rooting
procedures, so it is hard to generalize the results. They also did not
publish their datasets. In contrast, we not only consider current OS
versions but, more importantly, also provide a generic evaluation
procedure that can be repeated under different environmental
conditions.

Generally related to the present paper is a study by~\cite{son2013study} who describe
a tool for forensically sound Android data
acquisition using a custom recovery. While being interesting, this method
is mostly irrelevant on modern smartphones today due to the increased security standards.

\subsection{Paper Outline}

In Section~\ref{sec:zoo} we provide an overview of the different
backup mechanisms for Android and iOS. This is followed by detailed
information about our developed evaluation procedure, the data
classification strategy, and the device preparation in
Section~\ref{sec:method}.  The results of the evaluation, as well as
the device specific procedures, are presented in
Section~\ref{sec:android_eval} for Android and
Section~\ref{sec:ios_eval} for iOS, each accompanied by a comprehensive
analysis of the data.
In Section~\ref{sec:limit}, we lay out the limitations of our work
and draw a short conclusion in Section~\ref{sec:conclusion}.

\section{A Visit to the Zoo of Backup Methods} \label{sec:zoo}

The two major mobile platforms, Android and iOS, provide extensive
backup functionalities to secure critical data.
This data consists primarily of user- and application data, as well as
device settings. The backup options can be classified into different
categories according to their storage location, their scope and the type
of backed up data, as described in the following sections.
We will focus exclusively on the backup mechanisms provided by the OS while ignoring
third-party and vendor-specific solutions.

\subsection{Local vs.~Remote Backup}

Both platforms offer a possibility to store the backup data locally,
on a connected PC, or on the accompanying cloud services, i.e. Google Drive
for Android and iCloud for iOS.

The local backup mechanisms have a high value in forensics, due to
their ease of use and scope of data.
For Android, it can be executed with the \textit{backup} command
from the \textit{Android Debug Bridge (ADB)} command-line tool.
For iOS the local backup can be created natively on an Apple Mac,
using the iTunes software for Windows, or OS independent with
the open-source toolset \textit{libimobiledevice}~\citep{limd}, which is a community
project that reimplements Apple's proprietary iDevice communication protocol.

The remote backup is often created automatically, depending on the user's
settings, to either of the cloud services. Free storage is already
included for iCloud and Google Drive with a user's Google or Apple account.
Access to this data for forensic purposes is, however, complicated,
or in some instances impossible.

The trend of backups is moving strongly towards remote backups, with
the ADB backup command already marked deprecated since 2019.
Therefore, it might be omitted completely in future updates.
Furthermore, apps targeting Android 12  and up are automatically disabled
for ADB backup. On iOS, however, the local backup function is still actively supported.
iOS's local backup can additionally be decrypted using a user-defined password,
which leads to the inclusion of more sensitive user data in the backup.

\subsection{Full vs.~Selective Backup}

Every installed app can decide which of its files will be included or
excluded from the backup operation.
This can be done, for both Android and iOS by saving a file to a specific
filesystem location that is excluded or included by default or
by manually excluding or including files from the backup.
The platform's backup mechanism works mostly in an app or package oriented way,
which defines the granularity of the backup.
We refer to the backup of only one or more specified packages as
selective backup. Whereby, the backup, which iterates all packages,
is called a full backup.

However, there are exceptions, since not only app data can be included in a local backup.
On Android, the media folder, containing files like images or documents,
can optionally be included.
iOS follows the concept of domains in its backup mechanism, where one domain
in the backup data corresponds to a filesystem location and a certain group of files.
The package data is stored in an \emph{AppDomain} folder, one for each application
in the backup data.
Further domain examples are the \emph{CameraRollDomain}, which includes the
images from the phone's camera, or the \emph{HomeDomain}, with various files from
the user's home directory.

For Android, a selective backup can be created by specifying the packages
with ADB's backup command. By using the \texttt{-full} switch, all packages
are iterated and therefore included.
On iOS, no such possibility exists, only a full backup including all
domains and packages can be created.
The only influence on the amount of data included in an iOS backup is
by enabling or disabling the backup encryption, since encrypted backups contain
more sensitive data, like health data or stored passwords.

\subsection{File-based vs.~Content-based Backup}

\begin{figure}[htb]
    \centering
    \resizebox{0.49\textwidth}{!}{%
	\begin{tikzpicture}
    \def\spwidth{2.3}
    \def\spheight{4.5}
    \draw[rounded corners,line width=1.5mm] (0,0) rectangle (\spwidth,\spheight);
    \fill[rounded corners] (0,\spheight-0.4) |- (\spwidth*0.5,\spheight) -| (\spwidth,\spheight-0.4);
    \fill[rounded corners] (0,0.5) |- (\spwidth*0.5,0) -| (\spwidth,0.5);
    \draw[ultra thin,white] (\spwidth*0.8,\spheight-0.15) circle (0.1);
    \draw[rounded corners,very thin,white,fill=black!90] (\spwidth*0.4,0.05) rectangle (\spwidth*0.6,0.35);

    \def\corner{0.15in}
    \def\cornerradius{0.02in}
    \def\lwidth{0.02in}
    \def\h{0.9in}
    \def\w{0.6in}
    \def\iconmargin{0.1in}
    \def\topmargin{0.3in}
    \def\nline{10}
    \begin{scope}[yshift=4.2cm, xshift=0.7cm]
    \foreach[count=\i] \filename in {file 1, file 2, file 3}
    {
	\coordinate (nw) at ($(-0.05in*\i,-0.15in*\i)$);
	\coordinate (ne0) at ($(nw) + (\w, 0)$);
	\coordinate (ne1) at ($(ne0) - (\corner, 0)$);
	\coordinate (ne2) at ($(ne0) - (0, \corner)$);
	\coordinate (se) at ($(ne0) + (0, -\h)$);
	\filldraw [-, line width = \lwidth, fill=white] (nw) -- (ne1) -- (ne2)
	 [rounded corners=\cornerradius]--(se) -- (nw|-se) -- cycle;
	\draw [-, line width = \lwidth] (ne1) [rounded corners=\cornerradius]-- (ne1|-ne2) -- (ne2);
	\node [anchor=north west] at (nw) (f\i) {\small \ttfamily \filename};
	\foreach \k in {1,...,\nline}
	{
	  \draw [-, line width = \lwidth, line cap=round]
	    ($(nw|-se) + (\iconmargin,\iconmargin) + (0,{(\k-1)/(\nline-1)*(\h - \iconmargin - \topmargin)})$)
	      -- ++ ($(\w,0) - 2*(\iconmargin,0)$);
	}
    }
    \end{scope}

    \begin{scope}[yshift=4.2cm, xshift=4.3cm]
    \foreach[count=\i] \filename in {file 1, file 2}
    {
	\coordinate (nw) at ($(-0.05in*\i,-0.15in*\i)$);
	\coordinate (ne0) at ($(nw) + (\w, 0)$);
	\coordinate (ne1) at ($(ne0) - (\corner, 0)$);
	\coordinate (ne2) at ($(ne0) - (0, \corner)$);
	\coordinate (se) at ($(ne0) + (0, -\h)$);
	\filldraw [-, line width = \lwidth, fill=white] (nw) -- (ne1) -- (ne2)
	 [rounded corners=\cornerradius]--(se) -- (nw|-se) -- cycle;
	\draw [-, line width = \lwidth] (ne1) [rounded corners=\cornerradius]-- (ne1|-ne2) -- (ne2);
	\node [anchor=north west] at (nw) (f2\i) {\small \ttfamily \filename};
	\foreach \k in {1,...,\nline}
	{
	  \draw [-, line width = \lwidth, line cap=round]
	    ($(nw|-se) + (\iconmargin,\iconmargin) + (0,{(\k-1)/(\nline-1)*(\h - \iconmargin - \topmargin)})$)
	      -- ++ ($(\w,0) - 2*(\iconmargin,0)$);
	}
    }
    \end{scope}

    \begin{scope}[yshift=3.4cm, xshift=7cm]
    \foreach[count=\i] \filename in {file 3\\partial}
    {
	\coordinate (nw) at ($(-0.05in*\i,-0.15in*\i)$);
	\coordinate (ne0) at ($(nw) + (\w, 0)$);
	\coordinate (ne1) at ($(ne0) - (\corner, 0)$);
	\coordinate (ne2) at ($(ne0) - (0, \corner)$);
	\coordinate (se) at ($(ne0) + (0, -\h)$); 
	\filldraw [-, line width = \lwidth, fill=white] (nw) -- (ne1) -- (ne2)
	 [rounded corners=\cornerradius]--(se) -- (nw|-se) -- cycle;
	\draw [-, line width = \lwidth] (ne1) [rounded corners=\cornerradius]-- (ne1|-ne2) -- (ne2);
	\node [anchor=north west, text width=1cm] at (nw) (f3\i) {\small \ttfamily \filename};
	\foreach \k in {3,...,7}
	{
	  \draw [-, line width = \lwidth, line cap=round] 
	    ($(nw|-se) + (\iconmargin,\iconmargin) + (0,{(\k-1)/(\nline-1)*(\h - \iconmargin - \topmargin)})$)
	      -- ++ ($(\w,0) - 2*(\iconmargin,0)$);
	}
    }
    \end{scope}

    \draw[draw=blue!60!black!80!,-{Latex}] ($(f1.north east)+(0.2cm, -0.0cm)$) to [bend left=20]
	node[right, midway, text=darkgray] {} ($(f1.north east) + (2.8cm, +0.1cm)$);
    \draw[draw=blue!60!black!80!,-{Latex}] ($(f2.north east)+(0.2cm, -0.0cm)$) to [bend left=20]
	node[right, midway, text=darkgray] {} ($(f2.north east) + (2.8cm, +0.1cm)$);

    \node[draw=red, line width=0.5mm, minimum width=1.2cm, minimum height=0.72cm]
	at (1.09cm, 1.59cm) (cont1) {};
    \draw[draw=red!60!black!80!,-{Latex}] ($(cont1.south east)+(0cm, -0.1cm)$) to [bend right=20]
	node[right, midway, text=darkgray] {} ($(cont1.south east) + (5.4cm, -0.1cm)$);

    \node[font=\small, align=center] at ($(f22.north)+(0.2cm,1.2cm)$) (text1)
	{FILE-BASED ACQUISITION};
    \node[font=\small, align=center] at ($(f31.south)+(0.2cm,-2.3cm)$) (text2)
	{CONTENT-BASED ACQUISITION};
\end{tikzpicture}
    }
    \caption{Data types that can be acquired by the different acquisition methods.}
    \label{fig:eval_acq_types}
\end{figure}
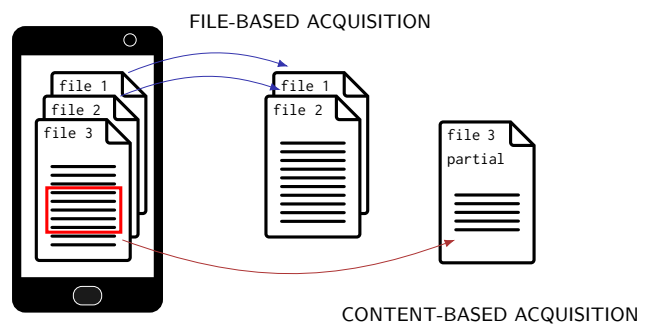

The problem with app-oriented backup design is that only the app itself
can really know the meaning of the data it contains, and therefore, which data
needs to be included in a backup.
The freedom of applications to determine the content of their backups results in two
different types of data that needs to be treated differently when
performing acquisition: \emph{file-based} and \emph{content-based}
data (see~\autoref{fig:eval_acq_types}).

We refer to one-by-one copies of files from the device's filesystem
as file-based data.
In contrast to file-based data, content-based data is a copy of only
partial data contents out of a file. The resulting file
may be of the same format with reduced or rearranged content or of a completely
different file format.  Therefore, the resulting data does not
represent an entire file but rather a subset of data stored in the
file (e.g. only one table from a bigger database).
There are content-based parts in both platforms' backup mechanisms, with
Android having a broader spectrum.
In Android, especially system packages use the so-called \emph{key-value backups} to store
valuable information, like device settings and the call history.
Key-value backup is a special backup mechanism, that stores the data as key-value pairs.
By default an app's backup is file-based, but a developer can choose to use
key-value backups instead, by defining its own \emph{BackupAgent} class or extending
from it. By doing so, the entire backup and restore procedure for the data
has to be implemented.
However, not many third-party apps are using this legacy mechanism anymore.
A special case of content-based data is Android's SMS backup, which
is not exactly a key-value backup but still only includes parts
from the original database file.

iOS has no equivalent to Android's key-value backups. Almost the entire backup
mechanism is file-based, with one exception which is the backup of the device's
\emph{keychain}. The keychain is a database
containing separately encrypted, sensitive bits of data, like the user's login
passwords. Parts of that keychain are included in an encrypted backup
as content-based data.

\subsection{App-downgrading}

The fact that the apps themselves decide on how to participate in a backup
implies that different versions of apps might also behave differently
regarding backup requests by the OS. If an app behaves more favorable
regarding backup in an older version than in the current version, one
strategy is to re-install the older version before performing the
backup operation. This procedure is known as \emph{app-downgrading}.
Selecting a particular version of an app therefore is a way to
influence the behavior of the local backup.

App-downgrading is of special importance for Android, since many
applications in their latest version no longer support local backups
of their data. However, in older app versions Android's local backup
mechanism was often available.
Therefore, by downgrading the app, it is possible to access local files
of many forensically relevant applications that would otherwise be inaccessible.
In Android the process involves multiple steps. First, a backup of the application's
current APK file is created, followed by uninstalling the current
version without deleting its local data, installing an older version
with backup support and executing the backup process for this
package. After a successful backup, the current app version can be reinstalled to
restore the initial state.

A patch was applied to the Android Open Source Project (AOSP) in 2016,
that should prevent data access by
downgrading~\citep{downgrade_patch}. However, this technique is still
working and is evidently also being used by forensic service
providers~\citep{oxy_down, cel_down, belka_down}.

\section{Evaluation Methodology and Experimental Setup} \label{sec:method}

There are different ways to conduct an evaluation, which can generally
be distinguished between more practical approaches, using real
devices, and laboratory approaches, with an emulator or development boards.
In this paper, the evaluation is performed with
actual hardware to simulate conditions as close to the
forensic practice as possible. But this approach also entails some
disadvantages, especially concerning the accessibility of data.

The focus of this work lies on (1) local backup and (2) the backup service
offered by the two platforms Android and iOS. We therefore neither
consider data acquisition from cloud storage nor from backups performed by
specialized third-party software. Within this scope, we evaluate data
acquisition both from file-based as well as content-based backup.

\subsection{Evaluation Model}
\label{sec:model}

Based on the previous considerations, we present a generic
practical evaluation approach, which serves as a template for the
specific evaluation processes. The method of \emph{differential forensic analysis},
similar to~\cite{garfinkel2012general}, is used as a baseline for the experiments.

To generalize from file- and content-based data, we
consider all backed up data as consisting of name-value pairs
$(n,v)$. For file-based backup, the name $n$ corresponds to the unique
filepath including the filename and the value $v$ to the file's content, i.e., the
complete bitstring stored under the filename in the filesystem. For
content-based backup, $n$ has to identify the unique file but also the data object
inside that file (e.g.,
a tag or column name from a database) while $v$ is the value of that
data object.

\begin{figure}[]
    \centering
    \resizebox{0.49\textwidth}{!}{%
	\begin{tikzpicture}
    \draw[draw=black,-{Latex}, font=\footnotesize] (0,0) to
	node[pos=0.93, above, text=darkgray] {time} (8,0);
    \draw[draw=red!60!black!80!,-, thick] (1,0.2) to
	node[below=0.2cm, text=darkgray, text width=2cm,align=center]
	{pre-acquisition\\reference data} (1,-0.2);
    \draw[draw=red!60!black!80!,-, thick] (4,0.2) to
	node[below=0.2cm, text=darkgray, text width=2cm,align=center]
	{perform\\backup} (4,-0.2);
    \draw[draw=red!60!black!80!,-, thick] (7,0.2) to
	node[below=0.2cm, text=darkgray, text width=2.5cm,align=center]
	{post-acquisition\\reference data} (7,-0.2);

    \draw[draw=black!40,-{Latex}] (1,-1.0) to
	node[below=0.2cm, text=black, font=\bfseries] {Pre} (1,-1.5);
    \draw[draw=black!40,-{Latex}] (4,-1.0) to
	node[below=0.2cm, text=black, font=\bfseries] {Backup} (4,-1.5);
    \draw[draw=black!40,-{Latex}] (7,-1.0) to
	node[below=0.2cm, text=black, font=\bfseries] {Post} (7,-1.5);
\end{tikzpicture}
    }
    \caption{Generic evaluation model}
    \label{fig:model}
\end{figure}
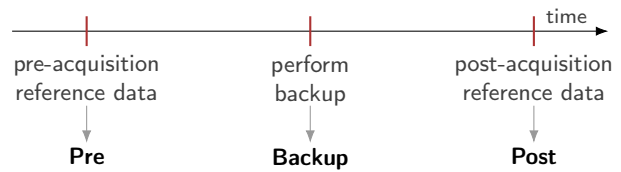

The evaluation is performed by systematically comparing the values
acquired from local backup to those values that were previously
existing under the same name on the device, a process that we call an
\emph{iteration} and which is depicted in~\autoref{fig:model}. To
reduce the effect of noise we perform multiple iterations.

In practice, the local backups contain both,
file-based and content-based data. For the file-based parts,
a simple hash-sum comparison is used to test for equality.
For the content-based data, which mostly stems from structured files,
like databases, the data itself is divided by the
smallest indivisible unit, according to the original filetype.
The comparison is made by simply comparing these values, according
to their original interpretation, which is mostly string-based.

In each iteration, three steps are performed:
\begin{enumerate}
\item We collect a set of \emph{pre-acquisition reference data} denoted
    as $\id{Pre}$. $\id{Pre}$ corresponds to
  the \emph{maximal set} of name-value pairs that can be obtained by the local
  backup method to be evaluated.

\item We then perform data acquisition using that backup method. The
  resulting set of name-value pairs is denoted by $\id{Backup}$.

\item After the backup is complete, we collect a set of
  \emph{post-acquisition reference data} denoted as $\id{Post}$. The
  scope of this data set should be identical to that of $\id{Pre}$.
\end{enumerate}

The necessity to create a maximal set results from the fact that the
selection of backup data is performed by each specific app, i.e., we
do not know beforehand which data objects will be part of $\id{Backup}$. In this
sense, $\id{Pre}$ and $\id{Post}$ are an overapproximation of $\id{Backup}$
regarding the set of name-value pairs. This implies that not all files
contained in $\id{Pre}$ and $\id{Post}$ will also be present in
$\id{Backup}$. Still, $\id{Backup}$ may contain data that is not present in either
$\id{Pre}$ or $\id{Post}$. We explain the different conditions that
may occur below.

The creation of $\id{Pre}$ and $\id{Post}$ critically depends on the
platform and the concrete device. Conceptually, it is similar to the
creation of $\id{Backup}$, but since it serves as a reference, any reliable data
acquisition method that creates a \emph{snapshot} of the filesystem
can be used here.

Creating snapshots on mobile devices requires privileged access (as root user).
The only possibility for full data access on iOS devices is to enable privileged access
by exploiting vulnerabilities.
This process is called \emph{jailbreaking}; it introduces many changes
to iOS and is accompanied by the installation of an unofficial app store most of the time.
But due to the lack of options, we chose this method for the iPhone,
to enable reference data access.

Certain Android phones offer the option to unlock their bootloader, which disables Android's
\emph{Verified Boot}, and thus, enables booting custom systems.
With an unlocked bootloader, there are two options to gain access to the filesystem data
for reference data acquisition.
A custom recovery system, like \emph{TWRP}, can be booted instead of Android and
can be used to access the phone's data or the Android OS can be modified to enable access
to the root account.
To be in line with iOS and to increase the methods comparability,
the reference data acquisition for Android will take place from within the OS as well,
by enabling root access on an unlocked phone.
This method was also chosen, because tests showed a huge amount of data alterations
during the reboot necessary for the recovery mode option, which can be mitigated with the
acquisition from within Android.

\subsection{Data Evaluation}

\begin{figure}[htb]
    \centering
    \resizebox{0.49\textwidth}{!}{%
	\begin{tikzpicture}
    \tikzset{
	box/.style={
	    rounded corners=0.05cm,
	    outer sep=0,%
	    inner sep=2pt,%
	    minimum width=1.0cm,%
	    minimum height=0.55cm,%
	    align=center}
    }

    \node[box, draw=white!60!red, font=\bfseries, text=darkgray, align=center]
	at (3.75,5) (E) {$E$};
    \node[box, draw=white!70!blue, font=\bfseries, text=darkgray, align=center]
	at (1.25,3.4) (N1) {$N_{over}$};
    \node[box, draw=white!70!blue, font=\bfseries, text=darkgray, align=center]
	at (2.5,3.4) (N2) {$N_{new}$};
    \node[box, draw=white!70!blue, font=\bfseries, text=darkgray, align=center]
	at (3.75,3.4) (N3) {$N_{both}$};
    \node[box, draw=white!70!green, font=\bfseries, text=darkgray, align=center]
	at (2.5,1.8) (V1) {$V_{eq}$};
    \node[box, draw=white!70!green, font=\bfseries, text=darkgray, align=center]
	at (3.75,1.8) (V2) {$V_{ch}$};
    \node[box, draw=black!30!, font=\bfseries, text=darkgray, align=center]
	at (0,0.2) (P1) {$P_{mis}$};
    \node[box, draw=black!30!, font=\bfseries, text=darkgray, align=center]
	at (1.25,0.2) (P2) {$P_{mback}$};
    \node[box, draw=black!30!, font=\bfseries, text=darkgray, align=center]
	at (2.5,0.2) (P3) {$P_{mpre}$};
    \node[box, draw=black!30!, font=\bfseries, text=darkgray, align=center]
	at (3.75,0.2) (P4) {$P_{nom}$};

    \draw[draw=black!30,-] (E.south) to (N1.north);
    \draw[draw=black!30,-] (E.south) to (N2.north);
    \draw[draw=black!30,-] (E.south) to (N3.north);

    \draw[draw=black!30,-] (N3.south) to (V1.north);
    \draw[draw=black!30,-] (N3.south) to (V2.north);

    \draw[draw=black!30,-] (V2.south) to (P1.north);
    \draw[draw=black!30,-] (V2.south) to (P2.north);
    \draw[draw=black!30,-] (V2.south) to (P3.north);
    \draw[draw=black!30,-] (V2.south) to (P4.north);

    \draw[draw=black!50,-, dotted] (-0.5,4.2) to
	node[pos=0.07, below, font=\scriptsize\color{black!60}] {union} (4.5,4.2);
    \draw[draw=black!50,-, dotted] (-0.5,2.6) to
	node[pos=0.07, below, font=\scriptsize\color{black!60}] {union} (4.5,2.6);
    \draw[draw=black!50,-, dotted] (-0.5,1.0) to
	node[pos=0.07, below, font=\scriptsize\color{black!60}] {union} (4.5,1.0);

    \draw [pen colour={black!50}, decorate, decoration = {calligraphic brace}]
	(4.7,5.45) -- node[right=0.2cm, text width=3.2cm]
	{Acquisition Experiment\\{\scriptsize\textcolor{black!50}{union of $Pre$ and $Backup$}}}
	(4.7,4.25);
    \draw [pen colour={black!50}, decorate, decoration = {calligraphic brace}]
	(4.7,4.15) -- node[right=0.2cm, text width=2.9cm]
	{Name Comparison\\{\scriptsize\textcolor{black!50}{between $Pre$ and $Backup$}}}
	(4.7,2.65);
    \draw [pen colour={black!50}, decorate, decoration = {calligraphic brace}]
	(4.7,2.55) -- node[right=0.2cm, text width=2.9cm]
	{Value Comparison\\{\scriptsize\textcolor{black!50}{between $Pre$ and $Backup$}}}
	(4.7,1.05);
    \draw [pen colour={black!50}, decorate, decoration = {calligraphic brace}]
	(4.7,0.95) -- node[right=0.2cm, text width=3.3cm]
	{Mismatch Classification\\{\scriptsize\textcolor{black!50}{using $Post$}}}
	(4.7,-0.55);

\end{tikzpicture}
    }
    \caption{Relations between the data sets}
    \label{fig:set_relations}
\end{figure}
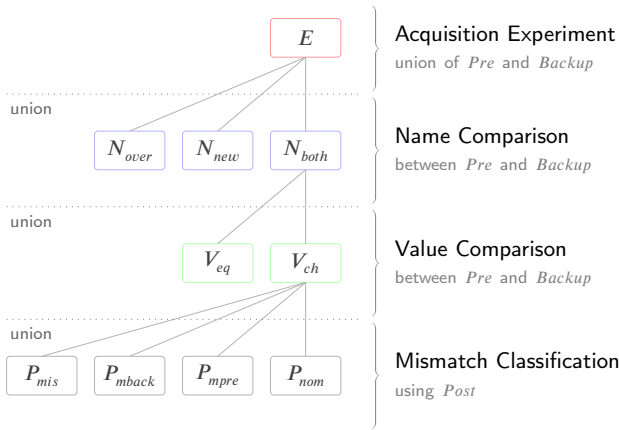

After performing the above steps, we evaluate the data. Generally, the
basic item of interest in our evaluation is a single name-value
pair. Depending on whether or not a pair with a specific name exists
in $\id{Pre}$, $\id{Post}$, or $\id{Backup}$ and what value is
associated with it, we distinguish the following data sets. For ease
of notation, by $\id{names}(X)$ we refer to the set of all names occurring
in the set $X$ of name-value pairs.
\begin{itemize}

\item $E$: The set of all observed names in $\id{Pre}$ together with
  those in $\id{Backup}$. Formally:
  $$\id{names}(\id{Pre}) \cup \id{names}(\id{Backup})$$
  This set characterizes the size of the data acquisition experiment.

\item $N_{over}$: The set of all names that are in $\id{Pre}$ but not in
  $\id{Backup}$. Formally:
  $$\id{names}(\id{Pre}) \setminus \id{names}(\id{Backup})$$
  This set characterizes the amount of existing data that was not
  backed up. Since $\id{Pre}$ is intentionally an overapproximation of
  $\id{Backup}$, the size of this set merely quantifies the degree of
  overapproximation.

\item $N_{new}$: The set of names that are in $\id{Backup}$ but not in
  $\id{Pre}$. Formally:
  $$\id{names}(\id{Backup}) \setminus \id{names}(\id{Pre})$$
  This set characterizes the spurious or ``new'' data in the
  backup that did not appear to exist before.

\item $N_{both}$: The set of all names that are both in $\id{Backup}$ and in
  $\id{Pre}$. Formally:
  $$\id{names}(\id{Pre}) \cap \id{names}(\id{Backup})$$
  This set characterizes the amount of existing data that was
  backed up.

  Note that by construction $|E| = |N_{over}| + |N_{new}| + |N_{both}|$.

\item $V_{eq}$: The set of all names of all name-value pairs where the
  names are both in $\id{Backup}$ and in $\id{Pre}$ and where the
  corresponding values are equal. Formally:
  $$ \id{names}( \id{Backup} \cap \id{Pre} ) $$
  Note that the intersection is on name-value pairs and not only on
  names. Therefore this set characterizes the elements that
  were backed up and whose content in the backup did not change.
 
\item $V_{ch}$: The set of all names of name-value pairs whose names are both in
  $\id{Pre}$ and $\id{Backup}$ but where the values are different. Formally:
  $$ \{ n : (n, v) \in\id{Pre} \land (n,v')\in\id{Backup}\land v\not= v' \} $$
  This set is comprised of backed up elements whose content
  appears to have changed.

  Note that by construction $|N_{both}| = |V_{eq}| + |V_{ch}|$

\end{itemize}
The following sets further subdivide the set $V_{ch}$, i.e., those data
objects that have changed before or during backup. Based on a
comparison with $\id{Post}$ this subdivision attempts to characterize
possible causes of the change of content:

\begin{itemize}

\item $P_{mis}$: The set of all names in $V_{ch}$ which do not occur in
  $\id{Post}$. Formally:
  $$ V_{ch} \setminus \id{names}(\id{Post}) $$

\item $P_{mback}$: The set of all names in $V_{ch}$ that appear in $\id{Post}$ and
  where the value in $\id{Backup}$ is equal to that in
  $\id{Post}$. Formally:
  $$ \{ n : n\in V_{ch} \land (n,v)\in\id{Backup} \land (n,v)\in\id{Post} \} $$
  This set characterizes those elements that changed before the backup
  but not thereafter. If the creation of $\id{Pre}$ happens shortly
  before the backup, elements in this set may have been changed \emph{by}
  the backup mechanism.

\item $P_{mpre}$: The set of all names in $V_{ch}$ where the value in
  $\id{Backup}$ neither matches the value in $\id{Pre}$ and
  $\id{Post}$ but where $\id{Pre}$ and $\id{Post}$ agree on
  the value. Formally:
  $$
  \begin{aligned}[t]
      \{ n : n\in V_{ch} &\land (n,v)\in\id{Backup} \land (n,v)\not\in\id{Post} \\
		    &\land (n,v') \in\id{Pre} \land (n,v'')\in\id{Post} \\
		    &\land v'=v'' \}
  \end{aligned}
  $$
  This set characterizes backed up elements, where the original data on the
  storage device does not appear to have changed but where the backed up
  value is different.

\item $P_{nom}$: The set of all names in $V_{ch}$ where the value in
  $\id{Backup}$ neither matches the value in $\id{Pre}$ and
  $\id{Post}$ and where also $\id{Pre}$ and $\id{Post}$
  disagree. Formally:
  $$
  \begin{aligned}[t]
      \{ n : n\in V_{ch} &\land (n,v)\in\id{Backup} \land (n,v)\not\in\id{Post} \\
		    &\land (n,v')\in\id{Pre} \land (n,v'')\in\id{Post} \\
		    &\land v'\not=v'' \}
  \end{aligned}
  $$
  This set characterizes elements where the cause of the mismatch is hard
  to determine because the data in $\id{Post}$ does not provide
  any helpful information.

  Note that by construction $|V_{ch}| = |P_{mis}| + |P_{mback}| + |P_{mpre}| + |P_{nom}|$.

\end{itemize}
The relations between the defined sets of data are depicted in~\autoref{fig:set_relations}.

For the items in $V_{ch}$, i.e., those that changed between $\id{Pre}$
and $\id{Backup}$, we provide an indication of the degree of changes
that occurred (or the degree of similarity).
Therefore, the similarity ratio $r$ is calculated with the
\textit{quick\_ratio} function from the SequenceMatcher in Python's difflib
library~\citep{py_difflib}, which can determine the similarity between
two byte sequences. This is done by comparing the values of the
two matching names for their likeness and as result, a value between $1.0$
(identical) and $0.0$ (completely different) is returned. The number of items considered for this value is the
cardinality of the set $V_{ch}$ and thus the sum of all
value-mismatches, with a name that exists in both $\id{Pre}$ and $\id{Backup}$.
To avoid the distortion caused by different sized sets, the average is
weighed by the sizes $s$.
Thus, the \textit{$\overline{r}_{w}$} value is the weighted
arithmetic mean of all $r$, calculated by the formula
$$
\overline{r}_{w} = \dfrac{1}{\sum_{i=1}^{|V_{ch}|}s_i}
\left( \sum\limits_{i=1}^{|V_{ch}|}s_i \cdot r_{i} \right)
$$
and similarly the weighted
standard deviation $\sigma_w$ is calculated
as
$$
\sigma_w = \sqrt{\dfrac{1}{\sum_{i=1}^{|V_{ch}|}s_i}
\left( \sum\limits_{i=1}^{|V_{ch}|} s_i (r_{i} - \overline{r}_{w})^2 \right)}.
$$
These values provide a general indication of the number of
changes within the files that are affected by a value-mismatch.

\subsection{Android Device Setup}

The device used for the Android evaluation is a Google Pixel 2. It has
the benefits of enabling a bootloader unlock and all released OS
images can be officially downloaded from Google's website.  Detailed
information about the device and the software tools
used for the acquisition are listed in~\autoref{tab:pixel2}.

\newcolumntype{F}{>{\centering\arraybackslash}m{3.8cm}}
\newcolumntype{H}{>{\itshape\arraybackslash}m{3.8cm}}
\begin{table}
    \centering
    \begin{tabular}{ H F }
	\toprule
	\multicolumn{2}{c}{\footnotesize\color{darkgray}\textbf{PIXEL 2 DEVICE INFORMATION}}\\
	\midrule
	\textit{Device} & Google Pixel 2 \\
	\textit{Codename} & walleye \\ 
	\textit{Storage} & 64 GB \\
	\textit{Android} & 11 \\
	\textit{Build Number} & RP1A.201005.004.A1 \\
	\textit{Encryption} & File-Based (FBE) \\
	\midrule
	\multicolumn{2}{c}{\footnotesize\color{darkgray}\textbf{SOFTWARE INFORMATION}} \\
	\midrule
	adb			 & 28.0.2-debian \\
	Android Backup Extractor & v20210530	\\
	Magisk & 25.2 \\
	\bottomrule
    \end{tabular}
    \caption{Hardware and software details of the Pixel 2 device and software information used for the acquisition.}%
    \label{tab:pixel2}
\end{table}

To perform the evaluation, a possibility to obtain reference data is
crucial. As described above, the reference data is
acquired within Android using root privileges. Furthermore, a
credible usage behavior has to be simulated. Accordingly, the device
needs to be prepared for our needs.

The first step was to unlock the bootloader of the phone to enable the
installation of custom partition images, which allows for the neccessary
system adjustments required for rooting.
The actual rooting process was accomplished with
\textit{Magisk}. Magisk is delivered as Android APK file which enables
patching the default \textit{boot.img} which is part of the factory
image. The patch enables apps and shell sessions over ADB to acquire
root privileges. Therefore, the patched \textit{boot.img} file has to
be flashed onto the device's \textit{boot} partition using
\textit{fastboot}.

After the device is set up to enable full data access, basic traces
to simulate a normal usage behavior have
to be created.  This includes some incoming and outgoing phone calls,
SMS messages, and emails.  Furthermore, contacts have been added and
pictures were taken with the device's camera.  Moreover, the widely
used messenger applications \textit{WhatsApp}, \textit{Telegram}, and
\textit{Facebook Messenger} were installed, alongside the
\textit{Firefox} browser, \textit{Instagram}, and \textit{Twitter}.

The state of the device before the acquisition is of importance as
well. To simulate an acquisition under realistic circumstances, the
device is assumed to be in a currently seized state with default
settings and the device is protected with a passcode.  Furthermore, in
compliance with the forensic guidelines for mobile
devices~\citep[p.28]{ayers2014guidelines}, the SIM card was removed
before the acquisition and it was ensured that the device cannot
connect to any network.

\subsection{iOS Device Setup}

The device chosen for the iOS backup evaluation is an iPhone 8, which was released
in 2017, but still supports the latest OS version (iOS 16). This device has the benefit of
being susceptible to the \textit{checkm8} BootROM vulnerability, and hence, privileged
access can be gained. Detailed device information and the software used for the acquisition
is provided in~\autoref{tab:iphone8}.

\begin{table}[htb]
    \begin{tabular}{H F}
	\toprule
	\multicolumn{2}{c}{\footnotesize\color{darkgray}\textbf{IPHONE 8 DEVICE INFORMATION}}\\
	\midrule
	\textit{Device} & iPhone 8 \\
	\textit{Model Number} & MQ6G2ZD/A \\
	\textit{Storage} & 64 GB \\
	\textit{iOS Version} & 14.6 \\
	\midrule
	\multicolumn{2}{c }{\footnotesize\color{darkgray}\textbf{SOFTWARE INFORMATION}} \\
	\midrule
	\textit{libimobiledevice} & 1.3.0-160	    \\
	\textit{iproxy} & 2.0.2-24 \\
	\textit{checkra1n} & 0.12.4 beta	    \\
	\textit{iOSbackup (python library)} & 0.9.923 \\
	\textit{irestore} & 1a78c2f \\
	\bottomrule
	\end{tabular}
    \caption{Hardware and software details of the iPhone 8 and software information used for the acquisition.}
    \label{tab:iphone8}
\end{table}

The iPhone must also be prepared for the evaluation with default usage behavior and privileged access
for reference data gathering. Therefore, the preparation steps include applying a jailbreak to
the device and creating usage traces.

The jailbreaking of the iPhone is accomplished with the \textit{checkra1n} tool~\citep{jb_cr} that uses
the checkm8 BootROM vulnerability.
After a successful jailbreak, a new app symbol with the checkra1n logo appears on the home
screen.
This app can be used to remove the jailbreak from the device or to install \emph{Cydia}, which
enables the installation of various packages and system modifications.
To enable access over SSH, Cydia is utilized to install the \textit{OpenSSH}
package. After the installation, an SSH connection over the default port 22 is
possible.

Similar to the Android device, some default usage behavior was simulated, to ensure
a realistic dataset. For that purpose, the apps \textit{WhatsApp}, \textit{Telegram},
\textit{Signal}, and \textit{Youtube} were installed, activated, and basic usage
was simulated. For the Apple account, the default settings for iCloud and privacy options
were used. Moreover, an email account was set up, SMS messages
were received and sent, contacts were created, calls simulated, and photos taken.

Again, the state of the device, in which the evaluation takes place needs to be considered.
Since the reference data is gathered from within the OS using the elevated privileges from
the jailbreak, the device's pre-acquisition state is powered on, locked, but without a lock
screen passcode, due to limitations of the jailbreak.
And in compliance with the aforementioned guidelines, all communication capabilities were disabled.

The method of reference data gathering must be implicitly trusted. Apart from the jailbreak tool,
of which the source code is only partially published, all tools are open-source and widely
used. However, the jailbreak tool is not explicitly used for data acquisition but only
provides the necessary privileges and its developers are well-known.
This leads to reasonable confidence about the soundness of the reference data acquisition
process.
However, since the jailbreak tool has full access to the OS, a compromised system cannot
entirely be ruled out, which should at least be noted.

\section{Android Backup Results} \label{sec:android_eval}

We now present the evaluation process and its results for Android using
ADB's local backup function. This is Android's most generic
acquisition method that can be used on any device, no matter the OEM
or software version.
Since the amount of data that can be acquired with a
local backup is limited, the method was complemented by app-downgrading,
with numerous apps that have special relevancy for forensics.

\subsection{Evaluation Process}

The data included in the backup process is a subset of the files from the
\path{/data/data/}, \path{/data/user_de/0/}, and
\path{/data/media/0/Android/data/} package directories, which depends
on the installed apps and their respective backup settings.  In
addition, the user's media files, which are located in the subfolders
of the \path{/data/media/} directory can be backed up, which is also
taken into account in the evaluation.

The reference data sets $Pre$ and $Post$ for the evaluation process are
acquired using the root account and the
\textit{tar} and \textit{nc} command-line tools. The reference data sets for the
full backup include the entire contents of the folders described above.
For the app-downgrading process, only the package-specific
subdirectory inside those folders was considered.

The actual backup acquisition process was executed by using ADB's
backup command. For a full backup, the corresponding ADB command is
used with the respective parameters to include all app data, media
files, and key-value data.
It results in an \textit{Android Backup (.ab)}
file, which is converted to a tar archive using the
\textit{Android Backup Extractor} utility.

For the evaluation of the app-downgrading method the apps displayed
in~\autoref{tab:downgrade_info} are chosen. These apps were activated and
used prior to the evaluation to ensure conditions and data sets are
reasonably realistic.
\begin{table*}[htb]
    \centering
    {
    \newcolumntype{F}{>{\centering\arraybackslash}m{4.7cm}}
    \newcolumntype{H}{>{\itshape\arraybackslash}m{3.2cm}}
    \newcolumntype{V}{>{\centering\itshape\arraybackslash}m{3.7cm}}
    \begin{tabular}{>{\color{black!70!white}\bfseries}H F V V }
	\toprule
	\multicolumn{1}{l}{\textbf{App}} &
	\multicolumn{1}{c}{\textbf{Package Name}} &
	\multicolumn{1}{c}{\textbf{Installed Version}} &
	\multicolumn{1}{c}{\textbf{Downgrade Version}} \\

	\arrayrulecolor{lightgray}
	\cmidrule(lr){1-1}
	\cmidrule(lr){2-2}
	\cmidrule(lr){3-3}
	\cmidrule(lr){4-4}
	\arrayrulecolor{black}
	WhatsApp & com.whatsapp & 2.23.1.76 & 2.11.399 \\
	Telegram & org.telegram.messenger & 8.9.3 & 1.3.1 \\
	Firefox & org.mozilla.firefox & 109.1.1 & 39 \\ 
	Facebook Messenger & com.facebook.orca & 393.0.0.18.92 & 14.0.0.16.14 \\
	Twitter & com.twitter.android & 9.73.0-release.0 & 4.0.3 \\
	Instagram & com.instagram.android & 267.0.0.18.93 & 7.9.0 \\

	\bottomrule
    \end{tabular}
}

    \caption{Information about the apps chosen for the downgrading acquisition.}
    \label{tab:downgrade_info}
\end{table*}
The versions to downgrade to were determined using a blog post by Oxygen forensics
as a reference~\citep{oxy_down}.
The respective APK files can easily be found with a quick online
search. This process, again, results in an Android Backup
file, which can be converted to a tar archive that contains the data
of the respective package.

The evaluation of file-based backup contents is based on comparing
filenames and hash-values of the file content.  However, since the file
structure of the backup does not match the structure of the
filesystem, the backup's file paths need to be mapped to their
original locations first. This is possible with the names of the
package's subfolders, which identify the original file location.

For the evaluation of content-based backup data, we focus on key-value
backups which can be utilized by applications instead of the more
modern file-based alternative. Furthermore, the backup of
SMS data is content-based because it is stored as a compressed
archive containing a JSON file, with partial data from the database.
The evaluation of key-value backups pose a challenge, because an app
has full control over the backed-up data and the format of the
key-value items can vary.  Therefore, analyzing the data from
third-party applications might not be useful, but is also not of
particular relevance, since none of the considered third-party
applications support this legacy backup mechanism.
However, Android system applications
still make use of this feature to backup valuable data.
\cite{kv_backups} analyzed key-value backups in a series of
blog posts, with a special focus on the call logs and system
settings, which are used as a reference for the evaluation.

The entire process was repeated for a total of 20 times for
the full backup and app-downgrading procedures to eliminate outliers.

\subsection{File-Based Backup}

The results of the file-based parts of the backup process are
presented in~\autoref{tab:res_back_andr}.  These include all
files that can be acquired with a local backup and
the app-downgrading process.  The vertical categories are therefore
divided by the \textbf{Full Backup} and \textbf{App-Downgrading} data,
which in turn is segmented by the individual applications (WhatsApp,
Telegram, Firefox, Facebook Messenger, Twitter, and Instagram).

Due to the large number of evaluation runs, the results shown in the table were grouped based
on equal values in the \textbf{Value Mismatch Classification} ($P_x$ classes). The number of
runs in a group can be taken from the first column (\#). Thus, all remaining values
in the table are averaged, as can be seen in the heading. The rows are arranged in descending
order according to the group's magnitude, i.e. according to its significance.

The \textbf{File Classification} ($N_x$) sets are an indicator of the validity of each
individual run.
Especially the $N_{new}$ class, which indicates omissions of $Backup$ data in the reference
set, is crucial for the value comparison.
Thus, an empty set, which we can observe in most cases, means that all files from
$Backup$ are included in the value comparison.
In the case of an empty $N_{new}$ set, the magnitude of set $N_{both}$ must be equal
to the size of $Backup$.
On some iterations, we encountered a non-empty $N_{new}$ set, which can occur due to
issues during the copying process of the reference data, or by newly
created temporary data in the timeslot between acquiring $Pre$ and $Backup$.

The \textbf{Value Classification} ($V_x$), provides an insight into the quality of the
acquired data from the backup process. Therefore, data alterations between $Pre$ and $Backup$
are reflected in the number of elements in set $V_{ch}$. As can be seen in the table,
the app-downgrading processes where affected in numerous cases, especially in the case
of Instagram.
However, since it is not a regular phenomenon, but only affects a limited number of runs,
it can be concluded that these changes are not an indication for
alterations due to the backup process or the app-downgrading procedure.
They merely capture the degree of concurrency of a given program.

According to the observed alterations, the interesting segments of the table
concerning the \textbf{Value Mismatch Classification} are the app-downgrading sections.
Here we can observe several classes of mismatches according to the $Post$ data comparison.
However, their occurrence is only sporadic and scattered among the classes. No
repeating patterns can be observed. Thereby, the suspicion of concurrent data changes,
during or between the individual acquisition processes, is reinforced.

\begin{table*}[ht]
    \centering
    {
\renewcommand{\arraystretch}{0.6}

\newcolumntype{A}{>{\scriptsize\color{black!50!white}\bfseries\centering\arraybackslash}m{0.2cm}}
\newcolumntype{B}{>{\scriptsize\centering\arraybackslash}m{0.6cm}}
\newcolumntype{C}{>{\scriptsize\centering\arraybackslash}m{0.6cm}}
\newcolumntype{D}{>{\scriptsize\centering\arraybackslash}m{0.45cm}}
\newcolumntype{E}{>{\scriptsize\centering\arraybackslash}m{0.9cm}}

    \arrayrulecolor{lightgray}
\begin{tabular}{A B B B |>{\bfseries}C | C C C | C C | D D D D | E E }
    \toprule
    \multicolumn{1}{c}{\scriptsize\textbf{}} &
    \multicolumn{3}{c}{\textbf{Filecount}} &
    \multicolumn{6}{c}{\textbf{File/Value Classification}} &
    \multicolumn{6}{c}{\textbf{Value Mismatch Classification}} \\
    \addlinespace[0.1cm]

    \arrayrulecolor{lightgray}
    \multicolumn{5}{c}{} &
    \multicolumn{3}{c}{\color{black!70!red}\bfseries\scriptsize $|E|$} &
    \multicolumn{2}{c}{\color{black!70!red}\bfseries\scriptsize $|N_{both}|$} &
    \multicolumn{4}{c}{\color{black!70!red}\bfseries\scriptsize $|V_{ch}|$} &
    \multicolumn{2}{c}{\color{black!70}\scriptsize Similarity Ratio} \\

    \cmidrule(lr){6-8}
    \cmidrule(lr){9-10}
    \cmidrule(lr){11-14}
    \cmidrule(lr){15-16} \\[-2.2ex]

    \multicolumn{1}{c}{\scriptsize \#}&
    \multicolumn{1}{c}{\scriptsize$\overline{|Pre|}$} & \multicolumn{1}{c}{\scriptsize $\overline{|Backup|}$} &
    \multicolumn{1}{c}{\scriptsize$\overline{|Post|}$} & 
    \multicolumn{1}{c}{\scriptsize$\overline{|E|}$} & \multicolumn{1}{c}{\scriptsize$\overline{|N_{over}|}$} &
    \multicolumn{1}{c}{\scriptsize$\overline{|N_{new}|}$} & \multicolumn{1}{c}{\scriptsize$\overline{|N_{both}|}$} &
    \multicolumn{1}{c}{\scriptsize$\overline{|V_{eq}|}$} & \multicolumn{1}{c}{\scriptsize$|V_{ch}|$} &
    \multicolumn{1}{c}{\scriptsize$|P_{mis}|$} & \multicolumn{1}{c}{\scriptsize$|P_{mback}|$} &
    \multicolumn{1}{c}{\scriptsize$|P_{mpre}|$} & \multicolumn{1}{c}{\scriptsize$|P_{nom}|$} &
    \multicolumn{1}{c}{\scriptsize$\overline{r}_{w}$} & \multicolumn{1}{c}{\scriptsize$\sigma_w$} \\

    \arrayrulecolor{black}
    \cmidrule(lr){2-4}
    \cmidrule(lr){5-10}
    \cmidrule(lr){11-16}

    \addlinespace[0.3em]
    \multicolumn{16}{c}{\color{black!60!white}\footnotesize\textbf{Full Backup}} \\
    \addlinespace[0.3em] 
    20 & 10853 & 1365 & 10856 & 10856 & 9511 & 0 & 1365 & 1365 & 0 & 0 & 0 & 0 & 0 & 0.0000 & 0.0000 \\
    \addlinespace[0.3em]
    \multicolumn{16}{c}{\color{black!60!white}\footnotesize\textbf{App Downgrading}} \\
    \addlinespace[0.1em]
    \multicolumn{16}{l}{\color{black!50!white}
    \scriptsize\textbf{WhatsApp}} \\
    \addlinespace[0.1em] 
    15 & 226 & 217 & 226 & 226 & 9 & 0 & 217 & 217 & 0 & 0 & 0 & 0 & 0 & 0.0000 & 0.0000 \\
    5 & 204 & 214 & 223 & 223 & 9 & 19 & 195 & 194 & 1 & 0 & 1 & 0 & 0 & 0.7951 & 0.0000 \\
    \addlinespace[0.3em]
    \multicolumn{16}{l}{\color{black!50!white}
    \scriptsize\textbf{Telegram}} \\
    \addlinespace[0.1em] 
    20 & 374 & 157 & 374 & 374 & 217 & 0 & 157 & 157 & 0 & 0 & 0 & 0 & 0 & 0.0000 & 0.0000 \\
    \addlinespace[0.3em]
    \multicolumn{16}{l}{\color{black!50!white}
    \scriptsize\textbf{Firefox}} \\
    \addlinespace[0.1em] 
    11 & 2179 & 185 & 2178 & 2179 & 1994 & 0 & 185 & 185 & 0 & 0 & 0 & 0 & 0 & 0.0000 & 0.0000 \\
    4 & 2178 & 185 & 2179  & 2179 & 1994 & 0 & 184 & 183 & 1 & 0 & 1 & 0 & 0 & 0.6936 & 0.0000 \\
    3 & 2179 & 185 & 2178  & 2179 & 1994 & 0 & 185 & 184 & 1 & 1 & 0 & 0 & 0 & 0.4442 & 0.0000 \\
    2 & 2178 & 185 & 2178  & 2179 & 1994 & 0 & 184 & 183 & 1 & 0 & 0 & 0 & 1 & 0.5565 & 0.0000 \\

    \addlinespace[0.3em]
    \multicolumn{16}{l}{\color{black!50!white}
    \scriptsize\textbf{Facebook Messenger}} \\
    \addlinespace[0.1em] 
    11 & 719 & 643 & 713 & 719 & 75 & 0  & 643 & 643 & 0 & 0 & 0 & 0 & 0 & 0.0000 & 0.0000 \\
    5 & 713 & 643 & 719  & 720 & 77 & 6  & 636 & 634 & 2 & 0 & 2 & 0 & 0 & 0.8648 & 0.0139 \\
    3 & 730 & 646 & 727  & 740 & 94 & 10 & 636 & 633 & 3 & 0 & 3 & 0 & 0 & 0.9915 & 0.0352 \\
    1 & 720 & 645 & 724  & 720 & 75 & 0  & 645 & 644 & 1 & 0 & 1 & 0 & 0 & 0.8550 & 0.0000 \\
    \addlinespace[0.3em]
    \multicolumn{16}{l}{\color{black!50!white}
    \scriptsize\textbf{Twitter}} \\
    \addlinespace[0.1em] 
    11 & 146 & 89 & 147 & 146 & 57 & 0 & 89 & 89 & 0 & 0 & 0 & 0 & 0 & 0.0000 & 0.0000 \\
    7 & 145 & 89 & 145  & 146 & 56 & 0 & 88 & 87 & 1 & 0 & 1 & 0 & 0 & 0.9615 & 0.0000  \\
    2 & 144 & 88 & 145  & 144 & 56 & 0 & 88 & 86 & 2 & 0 & 2 & 0 & 0 & 0.8160 & 0.0888  \\
    \addlinespace[0.3em]
    \multicolumn{16}{l}{\color{black!50!white}
    \scriptsize\textbf{Instagram}} \\
    \addlinespace[0.1em] 
    16 & 748 & 683 & 746 & 762 & 78 & 14 & 669 & 665 & 4 & 0 & 3 & 1 & 0 & 0.9942 & 0.0283\\
    1 & 741 & 683 & 748 & 741  & 58 & 0  & 683 & 682 & 1 & 0 & 1 & 0 & 0 & 0.9905 & 0.0000  \\
    1 & 763 & 712 & 776 & 774  & 62 & 11 & 701 & 698 & 3 & 0 & 2 & 1 & 0 & 0.9928 & 0.0325 \\
    1 & 745 & 685 & 750 & 763  & 78 & 18 & 667 & 661 & 6 & 0 & 5 & 1 & 0 & 0.9961 & 0.0223 \\
    1 & 744 & 684 & 749 & 762  & 78 & 18 & 666 & 659 & 7 & 0 & 6 & 1 & 0 & 0.9964 & 0.0242 \\

    \bottomrule
\end{tabular}
}

    \caption{Results of Android's file-based data evaluation of the full backup and the app-downgrading evaluation runs.}
    \label{tab:res_back_andr}
\end{table*}

\subsection{Content-Based Backup}

Now, the results and the scope of content-based data in
the backup process are presented. Therefore, the following
content-based files from the backup were considered for the evaluation
due to their importance:

\begin{itemize}[leftmargin=*, noitemsep, topsep=0pt]
    \item File: \textbf{\path{000000_sms_backup}}
\end{itemize}
The \path{com.android.providers.telephony/d_f/} directory from the backup includes this file,
it contains the user's SMS data and is available even without including key-value backup data.
There can be more than one SMS backup file, numbered sequentially in
ascending order, but due to the overseeable amount of messages, only one was created in
the process.
The $\id{Backup}$ data consists of the logical values of this file, mapped to their
original location. Those are the values \textit{address}, \textit{body}, \textit{date},
\textit{date\_sent}, \textit{read}, \textit{status}, and \textit{type} from the
\textit{sms} table and the \textit{recipients} from the
\textit{threads} and \textit{canonical\_addresses} table of the \textit{mmssms.db} file.
Therefore, the $\id{Pre}$ and $\id{Post}$ data sets were comprised of the
contents from the \textit{mmssms.db} database file, which again is an overapproximation
of the backup data.

\begin{itemize}[leftmargin=*, noitemsep, topsep=0pt]
    \item File: \textbf{\path{com.android.calllogbackup.data}}
\end{itemize}
This key-value backup from the \path{com.android.calllogbackup/k/} directory
contains information about the call history.
The $\id{Backup}$ data consists of a subset of the \textit{calllog.db} file.
This data includes the values
\textit{\_id}, \textit{number}, \textit{presentation}, \textit{date}, \textit{duration},
\textit{type}, \textit{subscription\_component\_name}, \textit{subscription\_id},
\textit{phone\_account\_address}, and \textit{block\_reason} from the \textit{calls}
table.
The database file, therefore, determines the content of the $\id{Pre}$ and
$\id{Post}$ data sets.

\begin{itemize}[leftmargin=*, noitemsep, topsep=0pt]
    \item File: \textbf{\path{com.android.providers.settings.data}}
\end{itemize}
This key-value backup file contains a subset of settings from various files
and is located in \path{com.android.providers.settings/k/}.
Therefore, $\id{Pre}$ and $\id{Post}$ were comprised of the values in the files
\textit{settings\_config.xml}, \textit{settings\_global.xml}, and
\textit{settings\_secure.xml}, which
contain various device settings, and of the values of \textit{WifiConfigStore.xml}, and
\textit{WifiConfigStoreSoftAp.xml}, that store various WiFi and network settings.

The detailed results of the content-based evaluation are depicted
in~\autoref{tab:res_back_andr_content}.
The content-based data was taken from the full backup runs used for the file-based
evaluation, whereby only one instance is shown in the table since there were no variations.
Furthermore, columns that do not provide additional information have been dropped.
The values with matching names were compared using a string-based comparison.
More key-value files existed in the backup data, but
most of them are dummy files or don't include relevant data, and since the content-based
evaluation is more time-consuming, only parts deemed forensically relevant were considered.

As can be seen in the table, the content-based parts of the evaluation process did not show
any signs of alterations or deviations over all backup iterations.
This result was expected, due to the very limited amount of data.

\begin{table}[htb]
    \centering
    {
\renewcommand{\arraystretch}{0.6}

\newcolumntype{B}{>{\scriptsize\centering\arraybackslash}m{0.75cm}}
\newcolumntype{C}{>{\scriptsize\centering\arraybackslash}m{0.5cm}}
\newcolumntype{D}{>{\scriptsize\centering\arraybackslash}m{0.5cm}}

    \arrayrulecolor{lightgray}
\begin{tabular}{B B B >{\bfseries}C C C C C }
    \toprule
    \multicolumn{3}{c}{\textbf{Atomic Elements}} &
    \multicolumn{5}{c}{\textbf{Name/Value Classification}} \\
    \addlinespace[0.1cm]

    \arrayrulecolor{lightgray}
    \multicolumn{4}{c}{} &
    \multicolumn{3}{c}{\color{black!70!red}\bfseries\scriptsize $|E|$} &
    \multicolumn{1}{c}{\color{black!70!red}\bfseries\scriptsize $|N_{both}|$} \\

    \cmidrule(lr){5-7}
    \cmidrule(lr){8-8}

    $|Pre|$ & $|Backup|$ & $|Post|$ & 
    $|E|$ & $|N_{over}|$ & $|N_{new}|$ & $|N_{both}|$ &
    $|V_{eq}|$ \\

    \arrayrulecolor{black}
    \cmidrule(lr){1-3}
    \cmidrule(lr){4-8}
    \addlinespace[0.1cm]
    \multicolumn{8}{c}{\color{black!50!white}\scriptsize\textbf{SMS Backup}} \\
    \addlinespace[0.1cm] 
365 & 56 & 365 & 365 & 309 & 0 & 56 & 56 \\
    \addlinespace[0.1cm]
    \multicolumn{8}{c}{\color{black!50!white}\scriptsize\textbf{Calllog Backup}} \\
    \addlinespace[0.1cm] 
101 & 20 & 101 & 101 & 81 & 0 & 20 & 20 \\
    \addlinespace[0.1cm]
    \multicolumn{8}{c}{\color{black!50!white}\scriptsize\textbf{Settings Backup}} \\
    \addlinespace[0.1cm] 
494 & 51 & 494 & 494 & 442 & 0 & 51 & 51 \\
\bottomrule
\end{tabular}
}

    \caption{Results of the content-based evaluation runs.}
    \label{tab:res_back_andr_content}
\end{table}

\subsection{Summary}

Overall, Android's backup process shows a high level of data integrity, which is only
affected by the program's and the OS's concurrency. However, these effects were
expected and should always be taken into account. Particularly in the case of the app
downgrading procedures, which require significant modifications, it is noteworthy that
the private application data of the corresponding packages does not experience any alterations
in the process. Of course, these results are not generally applicable and have to be
redone for different applications or under different circumstances.

\section{iOS Backup Results} \label{sec:ios_eval}

In the following, we present the results and the evaluation process of
iOS's backup function.
Contrary to Android, iOS's backup mechanism is still actively
supported and enables access to a large set of forensically important data.

\subsection{Evaluation Process}

The amount of data acquired by iOS's
backup mechanism cannot be specified in advance, since it depends on
the installed apps, their data quantities and backup policies, the
iCloud settings, and others.
Therefore, the set of data acquired as reference data for
$Pre$ and $Post$, is again an overapproximation, that ensures
a backing of all files contained in the backup. Since iOS's backup process also includes
files from various folders in its filesystem, the overapproximation will contain
the entire contents of the data partition, mounted at \path{/private/var/}.

The acquisition of $Pre$ and $Post$ is accomplished by using the previously
applied jailbreak, which enables root access over SSH.
The SSH access to the iPhone is established over USB
using \textit{iproxy} from the libimobiledevice project.
The data copy process is carried out by compressing the directory with the
\textit{tar} command whose output is piped over SSH to the connected PC.

For the backup creation, libimobiledevice's \textit{idevicebackup2}
tool is used since it enables easy utilization under Linux without
the need for iTunes or macOS.  Furthermore, two datasets, with and
without backup encryption are created.  The unencrypted backup
is comprised of a subset of the data of an encrypted backup, but
since no further processing steps are necessary, it can be ensured
that any modifications to the data originate from the backup process.
For the decryption of the encrypted backups, the python library \emph{iOSbackup}
is used. Because of some shortcomings of this library, which led to broken
database files, the \emph{irestore} tool, that is publicly available on
GitHub~\citep{keych_decr}, was used in addition.

iOS's backup contains file-based data for the most part. However, some
content-based data is present as well, which needs to be considered
separately.

Similar to the previous evaluation processes, iOS's local backup
mostly creates direct file copies. These files can then simply
be compared to their reference counterpart.  However, the assignment
of the files to their original filesystem location is not trivial
since all filenames and the file structure differ.  The mapping to
the respective filesystem data can be achieved with the contents of
the \textit{Manifest.db}, which contains a domain and the relative
path including the original filename for each file.

We now turn to the content-based data. On encrypted backups, some of the keychain entries
are included, which contain valuable user passwords and login information.
The decryption of this data is achieved by using the \textit{irestore} tool,
which offers the option to extract the keychain entries as a JSON file.
Since the keychain database in the reference data set is
encrypted using the device's hardware key, encryption of the database
entries is not possible. However, several login passwords are stored
on the phone, which can in turn be compared to the acquired backup
data.

A total of twenty evaluation runs for the encrypted- and unencrypted
backup processes are executed consecutively.  The chosen quantity
provides a good compromise between a big enough number of test runs for a
conclusive outcome while keeping the required storage space and the runtime
within reasonable limits.

\subsection{File-Based Backup}

The evaluation results of the file-based data of iOS's local backup are displayed
in~\autoref{tab:res_back_ios}.
The table is separated into \textit{Encrypted Backup} and \textit{Unencrypted Backup},
which each include an additional \textbf{$|V_{ch}|$ Overlapping Files} row, as
a measurement of files that are affected in each $P_x$ category in every run.
Similar to Android's result table, the rows are grouped by the $P_x$ categories
and sorted in descending order by the group's magnitude.

\begin{table*}[htb]
    \centering
    {
\renewcommand{\arraystretch}{0.6}
\newcolumntype{A}{>{\scriptsize\color{black!50!white}\bfseries\centering\arraybackslash}m{0.2cm}}
\newcolumntype{B}{>{\scriptsize\centering\arraybackslash}m{0.7cm}}
\newcolumntype{C}{>{\scriptsize\centering\arraybackslash}m{0.7cm}}
\newcolumntype{D}{>{\scriptsize\centering\arraybackslash}m{0.35cm}}
\newcolumntype{E}{>{\scriptsize\centering\arraybackslash}m{0.75cm}}
\centering
    \arrayrulecolor{lightgray}
\begin{tabular}{A B B B |>{\bfseries}C | C C C | C C | D D D D | E E }
    \toprule
    \multicolumn{1}{c}{\scriptsize\textbf{}} &
    \multicolumn{3}{c}{\textbf{Filecount}} &
    \multicolumn{6}{c}{\textbf{File/Value Classification}} &
    \multicolumn{6}{c}{\textbf{Value Mismatch Classification}} \\
    \addlinespace[0.1cm]

    \arrayrulecolor{lightgray}
    \multicolumn{1}{c}{\scriptsize} &
    \multicolumn{4}{c}{} &
    \multicolumn{3}{c}{\color{black!70!red}\bfseries\scriptsize $|E|$} &
    \multicolumn{2}{c}{\color{black!70!red}\bfseries\scriptsize $|N_{both}|$} &
    \multicolumn{4}{c}{\color{black!70!red}\bfseries\scriptsize $|V_{ch}|$} &
    \multicolumn{2}{c}{\color{black!70}\scriptsize Similarity Ratio} \\

    \cmidrule(lr){6-8}
    \cmidrule(lr){9-10}
    \cmidrule(lr){11-14}
    \cmidrule(lr){15-16} \\[-2.2ex]

    \multicolumn{1}{c}{\scriptsize \#}&
    \multicolumn{1}{c}{\scriptsize$\overline{|Pre|}$} &
    \multicolumn{1}{c}{\scriptsize $\overline{|Backup|}$} &
    \multicolumn{1}{c}{\scriptsize$\overline{|Post|}$} & 
    \multicolumn{1}{c}{\scriptsize$\overline{|E|}$} &
    \multicolumn{1}{c}{\scriptsize$\overline{|N_{over}|}$} &
    \multicolumn{1}{c}{\scriptsize$\overline{|N_{new}|}$} &
    \multicolumn{1}{c}{\scriptsize$\overline{|N_{both}|}$} &
    \multicolumn{1}{c}{\scriptsize$\overline{|V_{eq}|}$} &
    \multicolumn{1}{c}{\scriptsize$|V_{ch}|$} &
    \multicolumn{1}{c}{\scriptsize$|P_{mis}|$} & \multicolumn{1}{c}{\scriptsize$|P_{mback}|$} &
    \multicolumn{1}{c}{\scriptsize$|P_{mpre}|$} & \multicolumn{1}{c}{\scriptsize$|P_{nom}|$} &
    \multicolumn{1}{c}{\scriptsize$\overline{r}_{w}$} & \multicolumn{1}{c}{\scriptsize$\sigma_w$} \\

    \arrayrulecolor{black}
    \cmidrule(lr){2-4}
    \cmidrule(lr){5-10}
    \cmidrule(lr){11-16}
    \addlinespace[0.1em]
    \multicolumn{16}{c}{\color{black!60!white}\footnotesize\textbf{Encrypted Backup}} \\
    \addlinespace[0.3em] 
    14 & 39400 & 715 & 39401 & 39400 & 38685 & 0 & 715 & 630 & 85 & 0 & 1 & 84 & 0 & 0.9275 & 0.1438 \\
    4 & 39401 & 715 & 39399 & 39401 & 38686 & 0 & 715 & 629 & 86 & 0 & 2 & 84 & 0 & 0.9275 & 0.1438 \\
    1 & 39398 & 715 & 39398 & 39398 & 38683 & 0 & 715 & 628 & 87 & 0 & 2 & 84 & 1 & 0.9275 & 0.1438 \\
    1 & 39402 & 715 & 39402 & 39402 & 38687 & 0 & 715 & 628 & 87 & 0 & 3 & 84 & 0 & 0.9275 & 0.1438 \\
    \arrayrulecolor{lightgray}
    \hline
    \addlinespace[0.03cm]
    \multicolumn{10}{c|}{\scriptsize\color{black!70!white}$|V_{ch}|$ Overlapping Files} &\bfseries 0 &\bfseries 1 &\bfseries 84 &\bfseries 0 &&\\
    \addlinespace[0.03cm]
    \hline
    \arrayrulecolor{black}
    \addlinespace[0.3em]
    \multicolumn{16}{c}{\color{black!60!white}\footnotesize\textbf{Unencrypted Backup}} \\
    \addlinespace[0.3em] 
    18 & 39404 & 682 & 39404 & 39404 & 38722 & 0 & 682 & 611 & 71 & 0 & 1 & 70 & 0 & 0.9310 & 0.1351 \\
    1 &39405 & 682 & 39405 & 39405 & 38723 & 0 & 682 & 610 & 72 & 0 & 2 & 70 & 0 & 0.9310 & 0.1351 \\
    1 &39406 & 682 & 39407 & 39406 & 38724 & 0 & 682 & 609 & 73 & 0 & 3 & 70 & 0 & 0.9311 & 0.1351 \\

    \arrayrulecolor{lightgray}
    \hline
    \addlinespace[0.03cm]
    \multicolumn{10}{c|}{\scriptsize\color{black!70!white}$|V_{ch}|$ Overlapping Files} &\bfseries 0 &\bfseries 1 &\bfseries 70 &\bfseries 0 &&\\
    \addlinespace[0.03cm]
    \hline
    \arrayrulecolor{black}
\bottomrule
\end{tabular}
}

    \caption{Results of the file-based data evaluation for the encrypted and unencrypted backups.}
    \label{tab:res_back_ios}
\end{table*}

The general \textbf{Filecount} sections show that the amount of data in the
reference sets $Pre$ and $Post$ is slightly changing. This was expected due to the size
of the data sets and concurrently running programs.
However, the $Backup$ data stayed constant over all runs, which indicates a
high degree of determinism in this process.
The discrepancy between the amount of files included in encrypted and unencrypted backups
is based on the fact that encrypted backups generally contain more data.

The table's \textbf{File Classification} section especially highlights the validity of all runs,
since set $N_{new}$ is empty, and therefore, $N_{both}$ always contains the entire
$Backup$ set. This leads to the inclusion of all files in the value comparison
in every iteration.

The $P_{mis}$ set from the \textbf{Value Classification} category, that is empty in
each iteration, again reinforces the validity, as it indicates that all
the files from $N_{both}$ are also present in the $Post$ set, and therefore, can be
classified in greater detail.
$P_{mback}$ always contains at least one file, which is affected in every iteration,
as can be observed in the overlapping row. Since its values are always equal between $Backup$
and $Post$, the change occurs between $Pre$ acquisition and backup creation.
On further examination, the file contains unidentifiable metadata in XML format,
including a timestamp, that seems to be updated every time the backup process
is executed. The remaining occasionally occurring cases are probably attributable to
concurrency, which can also be stated about the single entry in $P_{nom}$, which has
a different value in all three data sets.
The most interesting observation is the high value of $P_{mpre}$ elements, that
are constant over all iterations, and as the overlapping row suggests, also concerns
the same files in each iteration.
This set represents files that are identical before and after the backup's execution,
but with value changes in the $Backup$ data. This suggests an alteration that is
happening during the backup process.
Furthermore, the \textbf{Similarity Ratio} section attests  high value similarities,
but the $\overline{r}_w$ and $\sigma_w$ values are almost identical over all runs. This leads
to the assumption that the files are subject to the same changes in every iteration.

On closer inspection, we found that the affected files were exclusively SQLite databases.
The majority of databases that were subject to changes were still contextually identical
to the reference set concerning their logical data, however, some showed alterations to various degrees.
This anomaly could be traced to the \textit{Write-Ahead Logging} mode of
an SQLite database, that, when in effect, creates a WAL file, which contains not yet
committed database changes.
These files, however, are not included in the backup process, but
when their content is committed manually to the databases in the reference data,
the mismatches disappear. This leads to the
conclusion, that the backup process commits the data of the WAL files to the backed up
databases.
According to~\citep{caithness2012forensic}, a deletion in a database with Write-Ahead Logging
that is not yet committed would still be recoverable. This information is therefore
missing in the acquired data of an iOS backup.

\subsection{Content-Based Backup}

The only content-based data of iOS's local backup is the keychain. The keychain on the
device consists of the single database file \path{/private/var/Keychains/keychain-2.db}.
Its contents, however, are separately decrypted using the device's hardware key, and therefore,
their decryption is not possible. Hence, no $Pre$ and $Post$ data sets are obtainable.
On a backup, the \textit{keychain-backup.plist} file is
created, which contains a subset of the database file as encrypted entries that
are only protected with the backup password.
The $Backup$ set is therefore comprised of the decrypted values from this file.
But since the contents of the \textit{keychain-2.db} is not decryptable, our evaluation
procedure cannot be applied to this data.
To at least provide an indication of the data's integrity, various browser login
information, as well as WLAN authentication data were stored on the device. This
data is secured in the keychain and included in the backup. The comparison of those
values to their $Backup$ counterpart showed no alterations.

\subsection{Summary}

In summary, iOS's backup process is a good source for forensically relevant data.
It should, however, be noted that the acquired data is partially subject to changes
due to the backup process.
But since the alterations stem from the inclusion of already executed database
changes, the data still reflects the actual database state.
Therefore, some forensically relevant data might be missing, but this does not invalidate
the acquired data.
Similar to Android, some concurrency related changes occured, which were expected
and cannot completely be ruled out since the acquisition process uses the OS of the mobile device.

\section{Limitations}
\label{sec:limit}

It is important to note that the differences between OS and application
versions, as well as device manufacturers, can be considerable.
Consequently, the result cannot be generalized, instead, the evaluation is intended to
provide a reference that must be redone under different environmental conditions.

Since backup data compatibility is necessary between different OS versions, we do
not expect the results of the general backup mechanism to differ significantly.
Changes can, however, not be ruled out without a thorough evaluation.
Especially app-downgrading should be taken with a grain of salt since the outcome
of that procedure is strongly dependent on the application and its handling of its
private data. According to a blog article by Oxygen Forensics~\citep{oxy_ad_limit},
huge differences between OS versions and device manufacturers exist for this method,
which might lead to the inability to restore the app's initial state or even the loss of data.
Accordingly, if this method is deemed necessary, an evaluation under similar environmental
conditions should be carried out first.

Furthermore, one can question the viability of local backups in the future, since, as already
mentioned, Android's local backup method has already been marked as deprecated.
However, the method is still widely used as an acquisition method and the evaluation of its
forensic suitability applies not only to future investigations but also serves as validation of
past cases.
If the ADB backend to create a local backup to a PC is ultimately removed from Android, it is
still
possible to copy the phone's data onto another device due to Android's data migration feature.
Accordingly, the backup data might still be obtainable using a prepared device, that serves as
an intermediate backup destination. However, the app-downgrading procedure might not be possible
anymore.
With iOS, on the other hand, it is unlikely for the local backup method to disappear in the
foreseeable future.
In general, some form of local backups will probably be around for quite some time, as
there are still users with limited internet access as well as those who prefer to handle their
private data locally due to privacy concerns.

\section{Conclusion and Future Work} \label{sec:conclusion}

Based on our evaluation model, including the data classification and categorization,
we were able to make comprehensive observations about the backup processes of Android and iOS.

In summary, our developed evaluation procedure was successfully
executed on two instances and provides, despite the described
limitations, a better understanding of the validity of the data
acquired during a backup acquisition.  Especially for forensic
purposes, where the data can determine the outcome of a legal case,
accuracy is of particular significance.

Since we cannot make our dataset publicly available due to privacy reasons, one of our
goals for future research is to repeat the experiments on a publishable yet realistic dataset.
Furthermore, the scope of the practical evaluation can always be extended to more OS
versions, device manufacturers, or apps.
Accordingly, a fully automated test environment with the use of virtual devices would
be a logical continuation.

\subsection*{Acknowledgements}

Work was supported by Deutsche
Forschungsgemeinschaft (DFG, German Research Foundation) as part of
the Research and Training Group 2475 ``Cybercrime and Forensic
Computing'' (grant number 393541319/GRK2475/1-2019).

We specifically wanted to thank Janine Schneider, Jan Gruber, Frank Breitinger and
Immanuel Lautner for their helpful comments and the useful discussion about the contents of
the paper.

\bibliography{backupevaluation}

\printcredits

\end{document}